\documentclass[12pt,aps,prd,preprint,tightenlines,superscriptaddress,
  showpacs,nofootinbib]{revtex4}
\newcommand{\PRE}[1]{{#1}} 

\usepackage{hyperref}      
\usepackage{bm}
\usepackage{epsfig}
\usepackage{color}
\usepackage{comment}
\usepackage{ulem}

\newcommand{\beqa}{\begin{eqnarray}}
\newcommand{\eeqa}{\end{eqnarray}}
\newcommand{\beq}{\begin{equation}}
\newcommand{\eeq}{\end{equation}}
\newcommand{\bay}{\begin{array}}
\newcommand{\eay}{\end{array}}

\newcommand{\ben}{\begin{enumerate}}
\newcommand{\een}{\end{enumerate}}
\newcommand{\bit}{\begin{itemize}}
\newcommand{\eit}{\end{itemize}}

\def\ord{{\cal O}}
\def\lt{\left}
\def\rt{\right}
\def\half{\frac{1}{2}}

\newcommand{\gweak}{g_{\text{weak}}}
\newcommand{\mweak}{m_{\text{weak}}}

\newcommand{\sigmaan}{\sigma_{\text{an}}}

\newcommand{\gev}{\text{GeV}}
\newcommand{\tev}{\text{TeV}}
\newcommand{\pb}{\text{pb}}

\newcommand{\m}{\text{m}}

\newcommand{\eqref}[1]{Eq.~(\ref{#1})}

\newcommand{\Eqref}[1]{Equation~(\ref{#1})}
\newcommand{\secref}[1]{Sec.~\ref{sec:#1}}

\newcommand{\figref}[1]{Fig.~\ref{fig:#1}}

\newcommand{\Figref}[1]{Figure~\ref{fig:#1}}
\newcommand{\tableref}[1]{Table~\ref{table:#1}}

\newcommand{\mgravitino}{M_{3/2}}

\newcommand{\neff}{N_{\text{eff}}}

\newcommand{\vev}[1]{\langle #1\rangle}

\renewcommand{\em}{\it}

\begin{document}

\preprint{UCI-TR-2011-24}

\title{ \PRE{\vspace*{0.5in}}
WIMPless Dark Matter from an AMSB Hidden Sector with No New Mass
Parameters
\PRE{\vspace*{0.3in}} }

\author{Jonathan L.~Feng}
\affiliation{Department of Physics and Astronomy, University of
California, Irvine, CA 92697, USA
\PRE{\vspace*{.2in}}
}

\author{Vikram Rentala}
\affiliation{Department of Physics, University of Arizona, Tucson, AZ
  85721, USA
\PRE{\vspace*{.5in}}
}
\affiliation{Department of Physics and Astronomy, University of
California, Irvine, CA 92697, USA
\PRE{\vspace*{.2in}}
}

\author{Ze'ev Surujon\PRE{\vspace*{.4in}}}
\affiliation{Department of Physics and Astronomy, University of
California, Irvine, CA 92697, USA
\PRE{\vspace*{.2in}}
}

\date{November 2011}
\PRE{\vspace*{0.6in}}

\begin{abstract}
\PRE{\vspace*{.3in}} We present a model with dark matter in an
anomaly-mediated supersymmetry breaking hidden sector with a
U(1)$\times$U(1) gauge symmetry.  The symmetries of the model
stabilize the dark matter and forbid the introduction of new mass
parameters.  As a result, the thermal relic density is completely
determined by the gravitino mass and dimensionless couplings.
Assuming non-hierarchical couplings, the thermal relic density is
$\Omega_X \sim 0.1$, independent of the dark matter's mass and
interaction strength, realizing the WIMPless miracle.  The model has
several striking features. For particle physics, stability of the dark
matter is completely consistent with $R$-parity violation in the
visible sector, with implications for superpartner collider
signatures; also the thermal relic's mass may be $\sim 10~\gev$ or
lighter, which is of interest given recent direct detection results.
Interesting astrophysical signatures are dark matter self-interactions
through a long-range force, and massless hidden photons and fermions
that contribute to the number of relativistic degrees of freedom at
BBN and CMB.  The latter are particularly interesting, given current
indications for extra degrees of freedom and near future results from
the Planck observatory.
\end{abstract}


\maketitle

\section{Introduction}
\label{sec:introduction}

The astrophysical evidence for dark matter is overwhelming, but the
mass and non-gravitational interactions of dark matter are unknown.
Under certain assumptions, however, one can place bounds on these
parameters.  One of the most interesting scales in high-energy physics
is the weak scale $v = 246~\gev$, which is currently being probed by
the Large Hadron Collider (LHC).  The framework of weakly-interacting
massive particle (WIMP) dark matter ties the mass and interaction
strength of a thermal relic dark matter particle to electroweak
physics. WIMPs, which are defined as particles with weak-scale masses
and couplings, naturally freeze out with the right relic density,
since
\begin{equation}
   \Omega_X \propto \frac{1}{\langle \sigmaan v \rangle} \sim
   \frac{\mweak^2}{\gweak^4} \ ,
   \label{eq:relicdensity}
\end{equation}
and for $\gweak \sim 0.6$ and $\mweak \sim v$, the thermal relic
density $\Omega_X$ is near the desired value $\Omega_{\text{DM}}
\approx 0.23$.  Since theories that explain the hierarchy problem
almost always introduce new weak-scale particles, they also typically
can include WIMP dark matter.

At the same time, \eqref{eq:relicdensity} implies that even particles
with different masses and couplings may have the right thermal relic
density, provided they have the same ratio $m/g^2$ as
WIMPs~\cite{Feng:2008ya,Feng:2008mu}.  As an example, such WIMPless
dark matter may arise in hidden sectors of gauge-mediated
supersymmetry (SUSY)-breaking models, provided that messengers
generate similar SUSY-breaking mass scales in the visible and hidden
sectors.  The possibility of dark matter with the correct thermal
relic density, but masses and couplings that differ, possibly
drastically, from WIMPs, opens up many new avenues for dark matter
detection~\cite{Feng:2008ya,Feng:2008mu,Feng:2008dz,Feng:2008qn,%
Kumar:2009ws,Barger:2010ng,Zhu:2011dz,McKeen:2009rm,Yeghiyan:2009xc,%
McKeen:2009ny,Badin:2010uh,Alwall:2010jc,Goodman:2010ku,Alwall:2011zm}.

Recently it has been shown~\cite{Feng:2011ik,Feng:2011uf} that models
with anomaly-mediated supersymmetry breaking
(AMSB)~\cite{Randall:1998uk,Giudice:1998xp} may also give rise to
WIMPless dark matter, without depending on messengers.  In AMSB,
superpartner masses scale as $m \sim (g^2/16\pi^2)\mgravitino$, where
$\mgravitino$ is the gravitino mass, a universal relation that holds
for all sectors, visible and hidden.  Hidden sectors, if they exist,
therefore generically have particles with the same ratio $m/g^2 \sim
\mgravitino/16\pi^2$ as WIMPs, and these particles are therefore
natural WIMPless candidates.

The visible sector in AMSB models enjoys the safety of being minimally
flavor violating~\cite{D'Ambrosio:2002ex,Buras:2000dm,%
Hall:1990ac,Chivukula:1987py}.  It is also highly predictive, as all
the new physics parameters are determined by the standard model (SM)
Yukawa and gauge couplings, along with three dimensionful parameters:
$\mgravitino$, $\mu$, and $B$. For example, gaugino masses are fixed,
relative to the gravitino mass, by the beta-functions to be
\begin{equation}
   M_1 : M_2 : M_3 : \mgravitino \approx 3.3 : 1 : -10 : 370 \ .
\end{equation}
Unfortunately, the AMSB framework also has problems: in its minimal
realization, sleptons are tachyonic, and the usual lightest
supersymmetric particle, the neutral Wino, has the right relic
abundance only for $m_{\widetilde W} \sim 3~\tev$, implying an
unnaturally large gluino mass $m_{\tilde g} \sim
30~\tev$~\cite{Giudice:1998xp}.

We will assume that the tachyonic slepton problem is solved, perhaps
by one of the mechanisms in the literature; see, for example,
Refs.~\cite{Pomarol:1999ie,Chacko:1999am,Katz:1999uw}.  As for the
second problem, since 30 TeV gluinos would reintroduce the hierarchy
problem, we may take it as a hint that the Wino is not a major
component of dark matter.  The Wino dark matter problem may be traced
back to the fact that SU(2) is nearly conformal in the minimal
supersymmetric standard model (MSSM), and so the Wino is
``accidentally'' light for its couplings.  In a hidden sector,
however, there is much more freedom in choosing gauge symmetries and
particle content.  We will take advantage of this and show that
WIMPless dark matter can originate from a U(1)$\times$U(1) hidden
sector.  Note that, in the models we present, the visible sector is
relieved from its duty to provide dark matter, and the hidden dark
matter particle is stabilized even without $R$-parity.  Dark matter
with a naturally correct thermal relic density is therefore perfectly
consistent with broken $R$-parity in this framework, with implications
for SUSY searches at colliders and elsewhere.

As noted above, WIMPless dark matter in AMSB has been explored in two
previous studies~\cite{Feng:2011ik,Feng:2011uf}.  Although these have
only scratched the surface of all model-building possibilities, it is
perhaps helpful to place this study in the context of the previous
two.  WIMPless dark matter requires that there be a bath of light
particles for the dark matter to annihilate to.  A natural possibility
is that this thermal bath is composed of massless gauge
bosons.\footnote{Goldstone bosons and chiral fermions are other
possibilities~\cite{Feng:2011ik}.}  It is, then, important that the
gauge symmetry not be broken (at least till freeze out).  In AMSB, the
generic expression for scalar soft masses is
\begin{equation}
   m_0^2\sim (y^4-y^2g^2-bg^4)\lt(\frac{\mgravitino}{16\pi^2}\rt)^2 \ ,
\label{genericsoft}
\end{equation}
where $y$ and $g$ denote Yukawa and gauge couplings, respectively, $b$
is the one-loop beta-function coefficient (with $b < 0$ for
asymptotically-free theories), and positive $\ord(1)$ coefficients in
front of each term have been suppressed.  In Ref.~\cite{Feng:2011ik},
asymptotically-free hidden sectors without Yukawa couplings were
considered.  Since $b<0$ for these sectors, $m_0^2 > 0$, and SUSY
breaking did not break the gauge symmetry. Provided the confinement
scale was sufficiently low, gauge bosons formed the thermal bath. In
Ref.~\cite{Feng:2011uf}, we considered Abelian models without Yukawa
couplings, where $b > 0$, but tachyonic scalars were avoided by
invoking $\mu$-terms to raises the scalar masses.  This led to some
extremely simple scenarios.  However, to realize the WIMPless miracle
in its purest form, these models required a mechanism to generate
$\mu$-terms of the same order as the SUSY-breaking parameters, as
discussed in Ref.~\cite{Feng:2011uf}.

In this paper we present another model with Abelian gauge symmetries,
but with masses completely determined by AMSB-induced soft
SUSY-breaking parameters.  Tachyonic scalars are avoided by
introducing Yukawa couplings, which raise the scalar masses and allow
us to construct a stable minimum for the scalar potential without
introducing a supersymmetric $\mu$-term by hand.  The model has a
U(1)$\times$U(1) gauge symmetry and 6 chiral superfields.  The
existence of a second U(1) (which is ultimately spontaneously broken)
and one more field compared to the models of Ref.~\cite{Feng:2011uf}
are needed to stabilize the potential without introducing
supersymmetric $\mu$-terms by hand. The other chiral fields are
required for anomaly cancellation.  Some of the particles, together
with the hidden photon, remain massless and contribute to the number
of extra degrees of freedom probed by Big Bang nucleosynthesis (BBN)
and the cosmic microwave background (CMB). Another prediction of the
model is that the dark matter candidate has long-range
self-interactions. Both the new massless degrees of freedom and the
self-interactions can be probed by current and future astrophysical
observations.

In the sections below, all particles and fields are in the hidden
sector unless otherwise noted, and we use MSSM-like notation for the
superfields and component fields.  For example, $\hat{e}$,
$\tilde{e}$, and $e$ denote a hidden electron superfield, selectron,
and electron, respectively, and $\hat{H}$, $H$, and $\tilde{H}$ denote
a hidden Higgs superfield, Higgs boson, and Higgsino.

\section{Model-Building Considerations}

The simplest Abelian model, supersymmetric QED (SQED), has the generic
problem of tachyonic sleptons in AMSB.  For concreteness, consider
SQED with one light flavor $(\hat e_+ , \hat e_-)$.  The positive
beta-function implies that the soft selectron mass parameters are
negative, breaking the U(1) spontaneously.  By itself, this is not
necessarily a problem, since the U(1) is hidden.  However, the
resulting quartic term in the potential,
\begin{equation}
   V_D=\frac{g^2}{2}\lt(\lt|\tilde{e}_+\rt|^2-\lt|\tilde{e}_-\rt|^2\rt)^2\ ,
\end{equation}
has a $D$-flat direction along
\begin{equation}
   \left| \tilde{e}_+ \right| =  \left| \tilde{e}_- \right| \ ,
\end{equation}
rendering the model unstable.

There are a few ways to stabilize the potential. First, supergravity
interactions would presumably stabilize the potential in any event.
However, if this is the dominant stabilizing effect, the scalars would acquire
vacuum expectation values (VEVs) at the Planck scale.  Whether such an
effect is parameterized by a hard SUSY-breaking quartic or by some
higher-dimensional operator, it would be related to Planck-scale
physics and therefore would not yield a viable WIMPless dark matter
candidate.

Another way to stabilize the potential is to introduce a
supersymmetric $\mu$-term by hand~\cite{Feng:2011uf}.  The obvious
drawback of this approach is that a new mass scale is being
introduced, thereby spoiling the natural WIMPless relation unless
there is a mechanism that generates it at the right scale, $\mu \sim
g^2 \mgravitino / (16\pi^2)$. The tachyon problem in SQED is therefore
transformed into a $\mu$-problem.  Note, however, the difference
between SQED and the MSSM: the former is a vector-like theory and
allows for $\mu$-terms for the sleptons.  In contrast, the MSSM lepton
sector is chiral, and requires extending the physical content of the
theory to solve the tachyonic slepton problem.

Here we will take a different approach that uses Yukawa interactions
in the hidden sector to stabilize the scalar potential.  Recall the
generic expression for scalar soft masses given in
\eqref{genericsoft}.  The presence of Yukawa interactions lifts the
scalar masses and may stabilize the potential. Of course, to allow
Yukawa interactions, the field content must be extended.

Perhaps the simplest extension of the SQED model above is obtained by
adding one gauge singlet superfield $\hat{H}$.  We may impose a
discrete $Z_3$ symmetry to avoid $\mu$-terms.  The most generic
renormalizable superpotential is then
\begin{equation}
   W=y \hat{H} \hat{e}_+ \hat{e}_- +  \frac{1}{6} \kappa \hat{H}^3 \ .
\end{equation}
Note that a non-zero value for $\kappa$ explicitly breaks the
(anomalous) global Peccei-Quinn (PQ) symmetry under which $\hat{H}$
has charge $2$ and $\hat{e}_+$ and $\hat{e}_-$ both have charge $-1$.
Including the new $F$-terms, the resulting scalar potential is
\begin{equation}
   V_{\rm SUSY}= \frac{g^2}{2}\lt(\lt|\tilde{e}_+\rt|^2-\lt|\tilde{e}_-\rt|^2\rt)^2 +
|y|^2\lt(|H|^2\lt|\tilde e_+\rt|^2+|H|^2\lt|\tilde e_-\rt|^2\rt)
           + \lt|\frac{1}{2} \kappa H^2 + y \tilde{e}_+ \tilde{e}_- \rt|^2 \ .
\end{equation}
The $D$-flat directions are lifted when $y$ and $\kappa$ are nonzero.
The soft SUSY-breaking parameters are
\begin{eqnarray}
   m_{\tilde\gamma} &=& \frac{g^2}{8\pi^2}\mgravitino \ , \nonumber\\
   m^2_{\tilde e_\pm} &=&\lt[-1-\lt(\frac{y}{g}\rt)^2
   +\frac{3}{4}\lt(\frac{y}{g}\rt)^4
   + \frac{1}{8}\lt(\frac{y}{g} \rt)^2\lt(\frac{\kappa}{g} \rt)^2
   \rt]
   m^2_{\tilde\gamma} \ , \nonumber\\
   m^2_{H} &=&\lt[-\lt(\frac{y}{g}\rt)^2
   +\frac{3}{4}\lt(\frac{y}{g}\rt)^4
    + \frac{1}{2} \lt(\frac{y}{g}\rt)^2\lt(\frac{\kappa}{g} \rt)^2
     +\frac{3}{16}\lt(\frac{\kappa}{g}\rt)^4
   \rt]m^2_{\tilde\gamma} \ , \nonumber\\
      A_{H \tilde{e}_+ \tilde{e}_-} &=& y \lt[2-\frac{3}{2} \lt(\frac{y}{g}\rt)^2
- \frac{1}{4}\lt(\frac{\kappa}{g}\rt)^2\rt] m_{\tilde\gamma} \ , \nonumber \\
       A_{HHH} &=& -\frac{3}{4} \kappa \lt[ 2 \lt(\frac{y}{g}\rt)^2
+ \lt(\frac{\kappa}{g}\rt)^2\rt] m_{\tilde\gamma} \ .
      \label{eq:qedspectrum}
\end{eqnarray}

Now that the $D$-flat directions are lifted, we examine this potential
for (meta-)stable minima.  For one of these vacua to have a WIMPless
dark matter candidate, it must satisfy several additional criteria:
\begin{itemize}
\item There should be at least one stable massive particle that plays
the role of dark matter.
\item There must be at least one light particle that serves as the
thermal bath.
\item The heavy dark matter particles must have tree-level
annihilations to the particles in the thermal bath to naturally get
the right relic density.
\end{itemize}

To examine the minima of the potential, we may begin by making various
assumptions for which fields acquire VEVs. Given one such assumption,
we then determine if there are ranges of the parameters $y/g$ and
$\kappa/g$ that give rise to stable minima with suitable WIMPless
candidates.  The possible symmetries that can prevent a heavy particle
from decaying into the thermal bath are electric charge, Lorentz
symmetry (the lightest fermion is stable), $R$-parity and, if $\kappa
\rightarrow 0$, the global PQ symmetry.  The particles that are
potentially light and can make up the thermal bath are the photon, the
electrons, the Higgsino, and, if U$(1)_{\text{PQ}}$ is a good symmetry
and is spontaneously broken, there may also be a light Goldstone boson
of the PQ symmetry.  However, in certain vacua, some (or all) of these
are massive.  Here are a few sample cases:
\begin{itemize}
\item None of the fields acquires a VEV: in this scenario, the photon,
the electron, and the Higgsino, are all massless.  However, none of
the massive particles is stable, since the decays $H \to e_+ e_-$, $\
\tilde{e}_{\pm} \to \tilde{H} \bar e_{\mp}$, and $\tilde{\gamma} \to
\tilde{H} e_+ e_-,$ are all allowed.  There is therefore no cold dark
matter candidate.
\item $H$ acquires a VEV, but $\tilde{e}_+$ and $\tilde{e}_-$ do not.
Note that this pattern of VEVs may be realized in some regions,
although \eqref{eq:qedspectrum} implies $m_H^2>m_{\tilde{e}_{\pm}}^2$.
In this case, the gauge symmetry is unbroken, so the photon is still
massless.  The fermions all become massive, and the lightest one is
stable.  Unfortunately, the model is constrained enough that the
lighter of the Higgsino and photino is always stable, since all its
decay modes are kinematically forbidden.  The Higgsino and photino do
not have tree-level annihilations to photons, and so would typically
overclose the universe.
\item $\tilde{e}_+$, $\tilde{e}_-$ and $H$ acquire VEVs. These VEVs
break the gauge symmetry. In general, the electrons and Higgsino will
be massive in these vacua. The only potential candidate for the
thermal bath is the pseudo-Goldstone boson of the PQ symmetry breaking
(in the $\kappa \rightarrow 0$ limit). This scenario merits further
study, but we note that the dark matter would annihilate through
derivative couplings, and therefore would not realize the WIMPless
miracle, at least in its purest form.
\end{itemize}
Although this simple Yukawa extension of SQED does not appear to
provide us with a WIMPless dark matter candidate, it illustrates many
of the potential problems and also suggests several ideas for model
building. In the next section, we will present a model that provides a
viable WIMPless dark matter candidate.

\section{A U(1)$\times$U(1) Model}
\label{sec:mirror}

Recall that \eqref{eq:qedspectrum} implies that the singlet extension
of the SQED model above satisfies the relation
$m_H^2>m_{\tilde{e}_{\pm}}^2$ everywhere throughout its parameter
space.  Although this by itself did not forbid the existence of vacua
with $\vev{H}\neq 0$ and $\vev{\tilde e_\pm}=0$, the constrained
nature of AMSB made it impossible to find a viable region without a
neutralino overabundance.  Therefore, we wish to modify the
singlet-added SQED model above so that $m_H^2<m_{\tilde{e}_{\pm}}^2$
can hold. One would hope that such a model would more easily realize
$\vev{H}\neq 0$ and $\vev{\tilde e_\pm}=0$ simultaneously.

To do this, we introduce a new U(1) gauge symmetry under which the
singlet is charged.  This gives rise to an additional negative
contribution to $m_H^2$.  We choose to gauge the PQ symmetry, namely
the U(1) that is ``axial'' with respect to the electron.  However, to
make the theory anomaly-free, we must introduce additional chiral
superfields.

Perhaps the simplest choice is a mirror duplicate sector with all the
charges inverted. This model has a U(1)$_A\times$U(1)$_B$ gauge
symmetry with gauge couplings $g_A$ and $g_B$, respectively.  We also
impose a $Z_3$ symmetry to forbid $\mu$-terms.  The field content and
the charges are given in \tableref{charges}.  Dark matter is
stabilized by hidden lepton flavor conservation.  $R$-parity ($R_p$)
is conserved, but it will play no role in stabilizing dark matter.  We
will use it only to distinguish between ``ordinary'' and
``superpartner'' fields.

\begin{table}[tb]
\begin{tabular}{|c | ccc | ccc|}
      \hline
      & \quad $\hat e_+$ \quad & \quad $\hat e_-$ \quad
      & \quad $\hat H_e$ \quad & \quad $\hat \mu_+$ \quad & \quad
      $\hat \mu_-$ \quad
      & \quad $\hat H_\mu$ \quad \\
      \hline
      U$(1)_A$ & 1 & $-1$ & 0 & $-1$ & 1 & 0\\
      U$(1)_B$ & 1 & 1 & $-2$ & $-1$ & $-1$ & 2\\
      $U(1)_e$ & 1 & $-1$ & 0 & 0 & 0 & 0\\
      $Z_3$ & 1 & 1 & 1 & 1 & 1 & 1\\
      $R_p$ & $-$ & $-$ & + & $-$ & $-$ & $+$ \\
      \hline
\end{tabular}
\caption{Superfields and their charges in the U(1)$\times$U(1) model.
\label{table:charges}}
\end{table}

The most generic superpotential is
\begin{equation}
   W = y_e\, \hat{H}_e \hat{e}_ + \hat{e}_-
     + y_\mu\, \hat{H}_\mu \hat{\mu}_+ \hat{\mu}_- \ .
\end{equation}
The model has four supersymmetric dimensionless couplings: $g_A$,
$g_B$, $y_e$, and $y_\mu$.  However, constraints from model building
and from the dark matter relic density will depend only on the three
ratios
\begin{equation}
   \tilde g_B \equiv \frac{g_B}{g_A} \ , \quad
   \tilde y_e \equiv \frac{y_e}{g_A} \ , \quad
   \tilde y_\mu \equiv \frac{y_\mu}{g_A} \ .
\end{equation}
Since annihilation of dark matter proceeds exclusively through
$A$-photon interactions, and so the annihilation cross section is
proportional to $g_A^4$, it is useful to express all the masses in
terms of $M_{\tilde{A}}$. The soft SUSY-breaking parameters induced by
AMSB are, then,
\begin{eqnarray}
   M_{\tilde{A}} &=& \frac{g_A^2}{4\pi^2}\mgravitino \ , \nonumber\\
   M_{\tilde{B}} &=&  3 \tilde g_B^2 M_{\tilde{A}} \ , \nonumber\\
   m^2_{\tilde e,\tilde\mu}
   &=&\lt(-\frac{1}{2}-\frac{1}{4}{\tilde y_{e,\mu}}^2
   +\frac{3}{16}\tilde y_{e,\mu}^4-\frac{3}{4}{\tilde y_{e,\mu}}^2{\tilde g_B}^2
   -\frac{3}{2}{\tilde g_B}^4\rt) M_{\tilde{A}}^2 \ ,\nonumber\\
   m^2_{H_{e,\mu}}&=&\lt(-\frac{1}{4}{\tilde y_{e,\mu}}^2
   +\frac{3}{16}\tilde y_{e,\mu}^4
   -\frac{3}{4}{\tilde y_{e,\mu}}^2{\tilde g_B}^2
   -6{\tilde g_B}^4\rt) M_{\tilde{A}}^2 \ ,\nonumber\\
   A_{e,\mu} &=& \tilde{y}_{e,\mu} \lt(1-\frac{3}{4}{\tilde y_{e,\mu}}^2
   +3{\tilde g_B}^2\rt) g_A M_{\tilde{A}} \ .
\end{eqnarray}

We are interested in solutions where at least one of the Higgs fields
acquires a VEV, but the selectrons and smuons do not.  In this case,
the $A$-photon remains massless and provides the thermal bath.  Note
that the relevant quartic term,
\begin{equation}
   V_{D_B} =\half g_B^2\lt(-2\lt|H_e\rt|^2+2|H_\mu|^2\rt)^2 \ ,
\end{equation}
has a $D$-flat direction along $\lt|H_e\rt|=\lt|H_\mu\rt|$.  To
maintain stability of the potential, the mass parameter along this
direction must therefore be positive, yielding the condition
\begin{equation}
   m^2_{H_e}+m^2_{H_\mu}>0 \ . \label{eq:flatdir}
\end{equation}
It follows that only one of the Higgs bosons can acquire a VEV.  Without
loss of generality, we choose this field to be $H_e$.  Minimizing the
potential results in
\begin{equation}
   \vev{H_e}^2=\frac{-m^2_{H_e}}{4g_B^2} \ .
\end{equation}
This VEV generates masses for the electrons and the $B$-gauge boson,
and it contributes to the masses of the selectrons and neutralinos.

In the bosonic sector, the physical Higgs and the $B$-gauge boson both
acquire the same mass,
\begin{equation}
   m^2_{H_e^0}=M^2_B=-2m^2_{H_e} \ .
\end{equation}
The selectron and smuon masses are
\begin{eqnarray}
   \Delta V = &&
   \pmatrix{\tilde e_+ & \tilde e_-^*}
   \pmatrix{m^2_{\tilde e_+}\! - \! 2g_B^2\vev{H_e}^2+\lt|y_e\rt|^2\vev{H_e}^2 &
   A_e\vev{H_e}\cr
               A_e^*\vev{H_e} & m^2_{\tilde e_-} \! - \! 2g_B^2\vev{H_e}^2
               +\lt|y_e\rt|^2\vev{H_e}^2}
   \pmatrix{\tilde e_+^* \cr \tilde e_-}\nonumber\\
   &&+\lt(m^2_{\tilde\mu_+}+2g_B^2\vev{H_e}^2\rt)\lt|\tilde\mu_+\rt|^2
   +\lt(m^2_{\tilde\mu_-}+2g_B^2\vev{H_e}^2\rt)\lt|\tilde\mu_-\rt|^2 \
   .
\end{eqnarray}
The resulting mass eigenvalues of the selectrons are
\begin{eqnarray}
   m^2_{\tilde e_2} &=& m^2_{\tilde e}-2g_B^2\vev{H_e}^2+|y_e|^2\vev{H_e}^2
   + |A_e|\vev{H_e} \ , \nonumber \\
   m^2_{\tilde e_1} &=& m^2_{\tilde e}-2g_B^2\vev{H_e}^2+|y_e|^2\vev{H_e}^2
   - |A_e|\vev{H_e} \ .
\end{eqnarray}
Note that we have used $m^2_{\tilde e} = m^2_{\tilde e_+} =
m^2_{\tilde e_-}$.  The singlet $H_\mu$ acquires a negative
contribution to its mass from the $D$-term, such that its physical
mass is
\begin{equation}
   \lt(m^{\text{phys}}_{H_\mu}\rt)^2 = m_{H_\mu}^2-4g_B^2\vev{H_e}^2
   =m_{H_e}^2+m_{H_\mu}^2 \ .
\end{equation}
We see that requiring the $D$-flat direction to be stable,
\eqref{eq:flatdir}, is equivalent to requiring
$\lt(m^{\text{phys}}_{H_\mu}\rt)^2 > 0$, as expected.

In the fermionic sector, $e_+$ and $e_-$ combine into one Dirac
fermion, the electron $e$, with mass $m_e = y_e \vev{H_e}$.  The muons
are massless and form part of the thermal bath.  There are four
neutralinos in the model: $\tilde A$ and two combinations of $\tilde
B$ and $\tilde H_e$ are massive, but $\tilde H_\mu$ is massless and is
part of the thermal bath.

\begin{figure}
\centering
\includegraphics[width=0.495\textwidth]{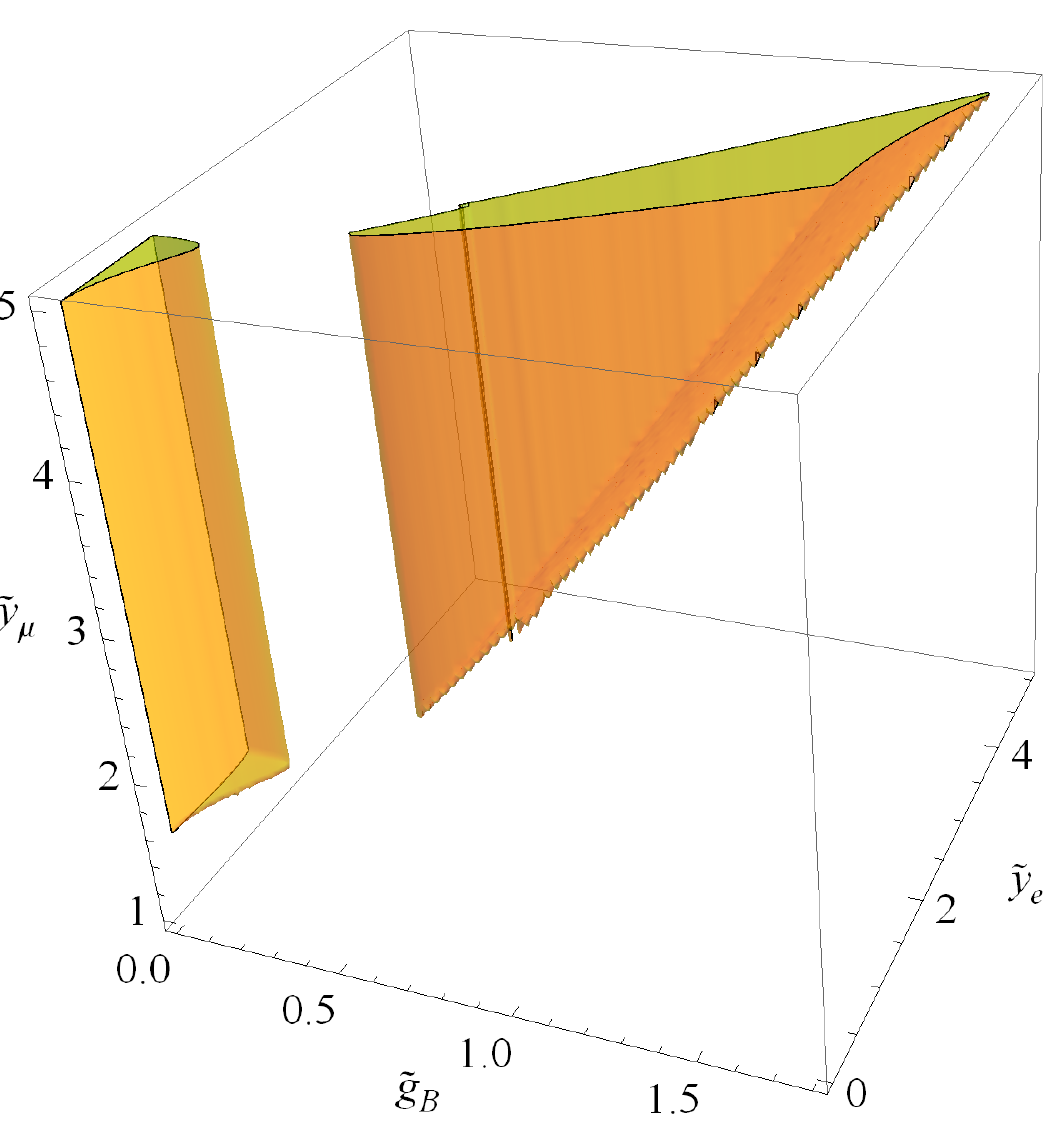} \hfil
\includegraphics[width=0.495\textwidth]{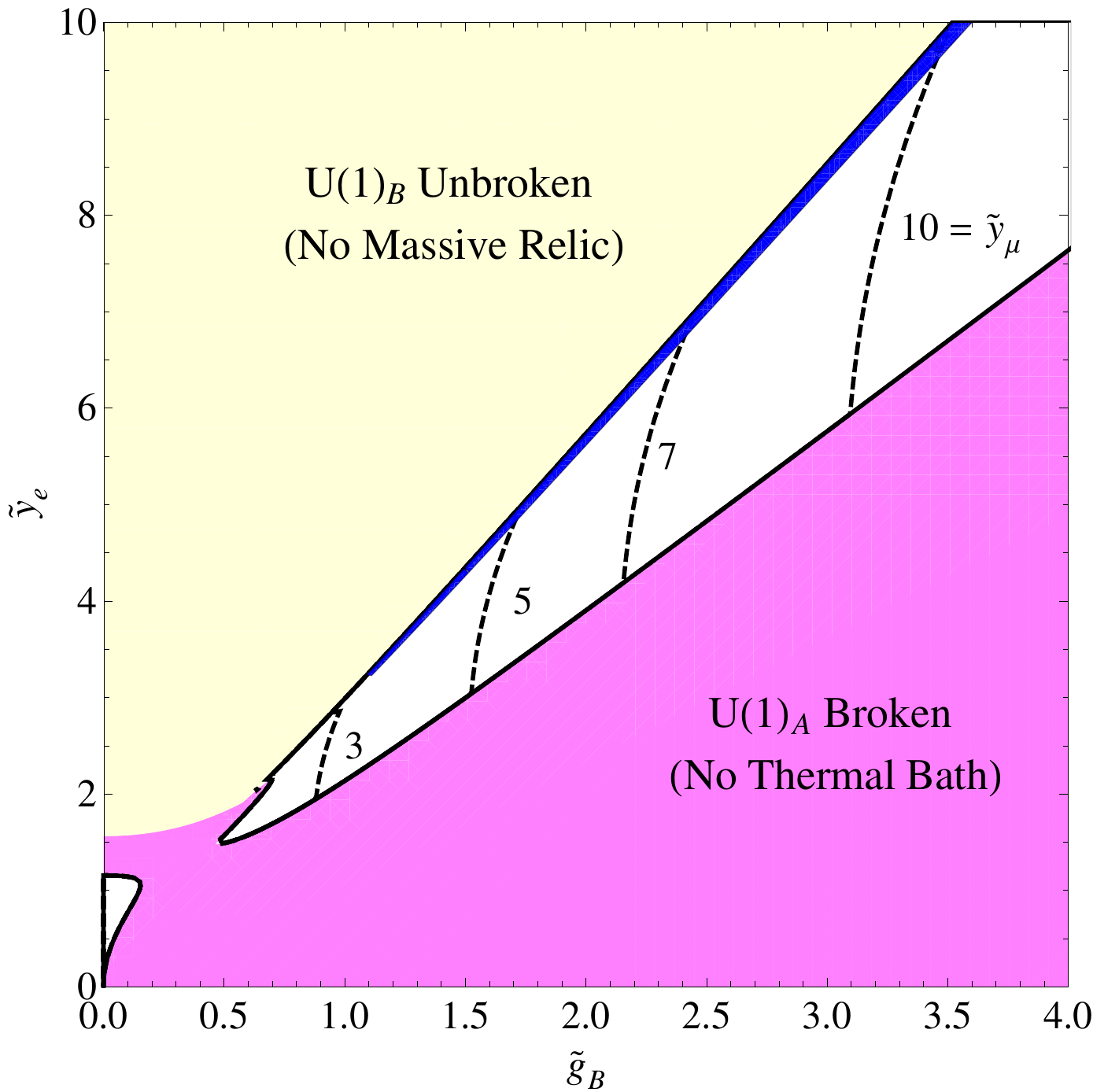}
\caption{{\em Left:} Allowed regions in the $(\tilde g_B, \tilde{y}_e,
\tilde{y}_\mu)$ parameter space of the U(1)$_A \times$U(1)$_B$ model.
{\em Right:} A projection of the allowed parameter space onto the
$(\tilde{g}_B, \tilde{y}_e)$ plane.  The light yellow and medium
magenta shaded regions are excluded for the reasons indicated.  Dark
matter is composed of selectrons everywhere in the viable region,
except inside the dark blue band, where it is electrons.  At
tree-level, the mass of the dark matter particle is independent of
$\tilde{y}_\mu$, as long as $\tilde{y}_\mu$ is in a viable region of
parameter space.  Contours of minimum $\tilde{y}_\mu$ for given values
of $(\tilde{g}_B, \tilde{y}_e)$ are shown.  Regions to the right of
the $\tilde{y}_\mu=\text{const.}$ curves are not viable for
$\tilde{y}_\mu > \text{const.}$, since the constraint
$m_{H_e}^2+m_{H_\mu}^2>0$ cannot hold, and the potential is unstable.
\label{fig:paramspace}}
\end{figure}

The rough picture of the spectrum is therefore:
\begin{itemize}
\item Massive particles:
1 $B$-gauge field,
1 physical Higgs ($H_e$),
1 Dirac electron ($e$),
3 heavy neutralinos ($\tilde A$, $\tilde B, \tilde H_e$), and
5 complex scalars ($H_{\mu}$, $\tilde e_{1,2}$, $\tilde \mu_\pm$).
\item Massless particles:
1 $A$-photon,
1 Higgsino ($\tilde H_\mu$), and
2 Weyl muons ($\mu_\pm$).
\end{itemize}
The potential candidates for dark matter are either the electron or
the lighter selectron $\tilde{e}_1$, with the lighter of these being
stabilized by an accidental global U(1) symmetry analogous to lepton
flavor. Note that the mass of the dark matter particle is independent
of $\tilde{y}_\mu$, as long as $\tilde{y}_\mu$ is in a viable region
of parameter space, as can be seen in \figref{paramspace}. (A weak
dependence will appear once higher-order corrections are included.)
All the other massive particles decay to a combination of the dark
matter particle and the massless fields.  The various decay channels
are listed in \tableref{decays}.

\begin{table}[tb]
\begin{tabular}{c c c}
      \hline\hline
      Particle & & Sample Decay Channel\\
      \hline\hline
      Heavy gauge boson $B$ & & $\mu_+\mu_-$ \\
\hline
      electron-type Higgs $H_e$ & & $A A$ \\
      \hline
      Neutralinos ($\tilde A$, $\tilde B$, $\tilde H_e$) & &
      $\mu_+\mu_-\tilde H_\mu$ \\
      \hline
      Muon-type Higgs $H_\mu$ & & $AA$, $\mu_+\mu_-$\\
      \hline
      Smuons $\tilde\mu_\pm$ & & $\mu_\pm\tilde H_\mu$\\
      \hline
      Heavy selectron $\tilde e_2$ & & $\tilde e_1 A$\\
      \hline\hline
\end{tabular}
\caption{Various decay channels for the heavy fields.  If
$m_{\tilde{e}_1} > m_e$, the lighter selectron decays through
$\tilde{e}_1 \to e \tilde{A}^{(*)}$, and if $m_e > m_{\tilde{e}_1}$,
the electron decays through $e \to \tilde{e}_1 \tilde{A}^{(*)}$, but
the lighter of $\tilde{e}_1$ and $e$ is stable and forms dark matter.
\label{table:decays}}
\end{table}

\Figref{paramspace} shows the viable regions in the $(\tilde g_B,
\tilde{y}_e, \tilde{y}_\mu)$ parameter space, namely those regions
where U$(1)_A$ is not broken (selectrons/smuons do not acquire a VEV,
and massless photons provide the thermal bath), U$(1)_B$ is broken
($H_e$ acquires a VEV, providing mass for the electrons), and the
potential along the $D$-flat direction is stabilized
($m_{H_e}^2+m_{H_\mu}^2>0$).  Although most of the viable region
admits scalar dark matter ($\tilde{e}_1$), dark matter is made of
fermions ($e$) in the narrow dark blue band.  This region has a small
Higgs VEV $\vev{H_e}$, and thus the electron is lighter than the
selectrons.  At another boundary of the scalar dark matter region the
scalars become massless.  Beyond that boundary, U$(1)_A$ is
spontaneously broken and there is no viable WIMPless dark matter.

\section{Relic Density}
\label{sec:relicdensity}

The thermal relic density of a dark matter particle $X$ annihilating
via $S$-wave processes is given by~\cite{Feng:2011uf} (see
Refs.~\cite{Feng:2008mu,Das:2010ts,Feng:2011ik} for a general
treatment)
\begin{equation}
\Omega_X \approx \xi_f \, \frac{0.17~\pb}{\sigma_0}
\simeq 0.23 \ \xi_f \, \frac{1}{k_X}
\left( \frac{0.025}{\alpha_{X}} \frac{m_X}{\tev} \right)^2 \ ,
\end{equation}
where $k_X$ is an $\ord(1)$ constant defined by $\sigma_0 \equiv k_X
\pi \alpha_X^2 / m_X^2$, $\alpha_X\equiv g_X^2/(4\pi)$ is the coupling
related to the annihilation process, and $\xi_f \equiv
T_f^{\text{h}}/T_f^{\text{v}}$ is the ratio of the hidden to visible
sector temperatures when the hidden dark matter freezes out.

\begin{figure}
  \centering
    \includegraphics[width=0.486\textwidth]{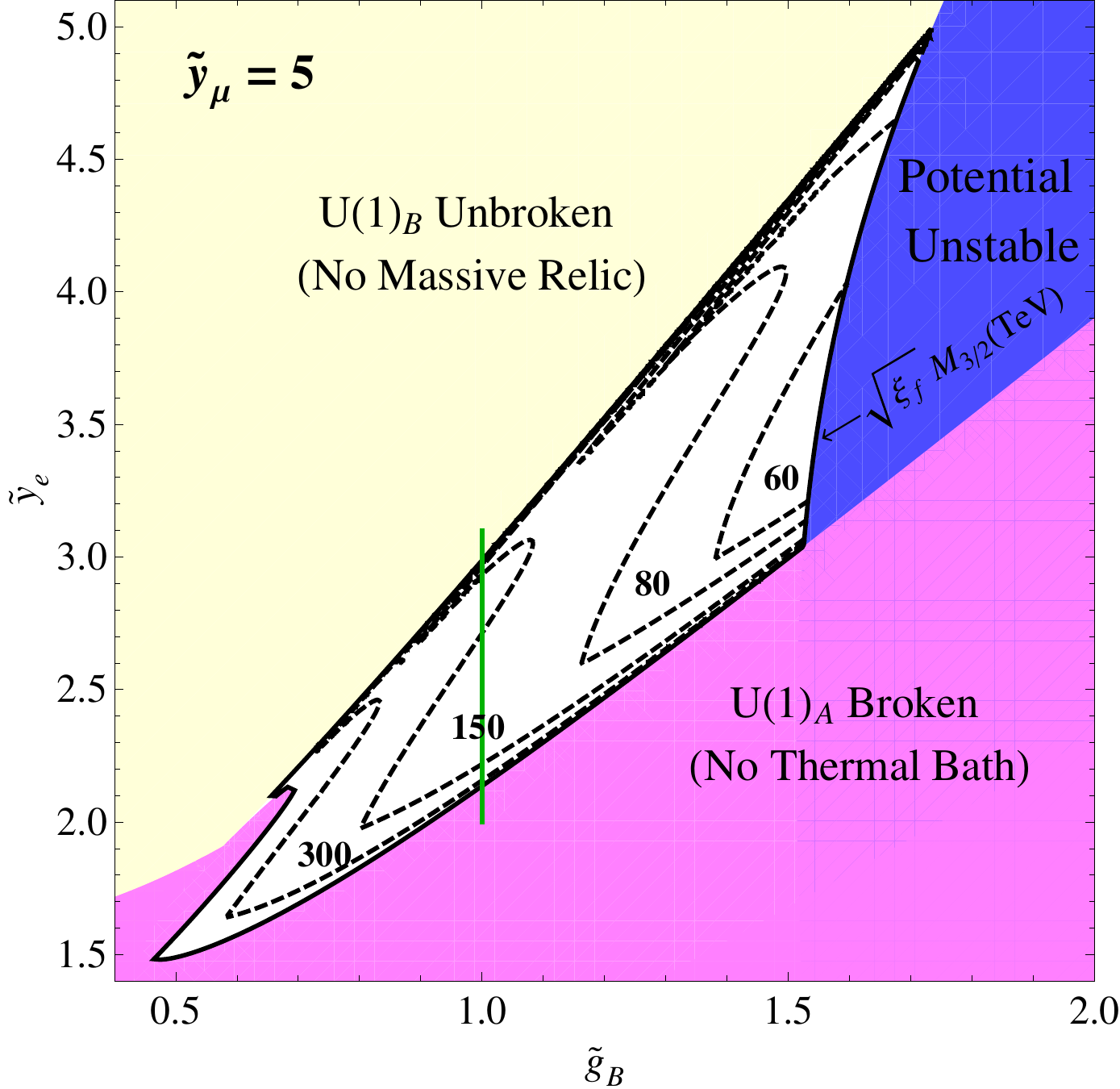} \hfil
    \includegraphics[width=0.494\textwidth]{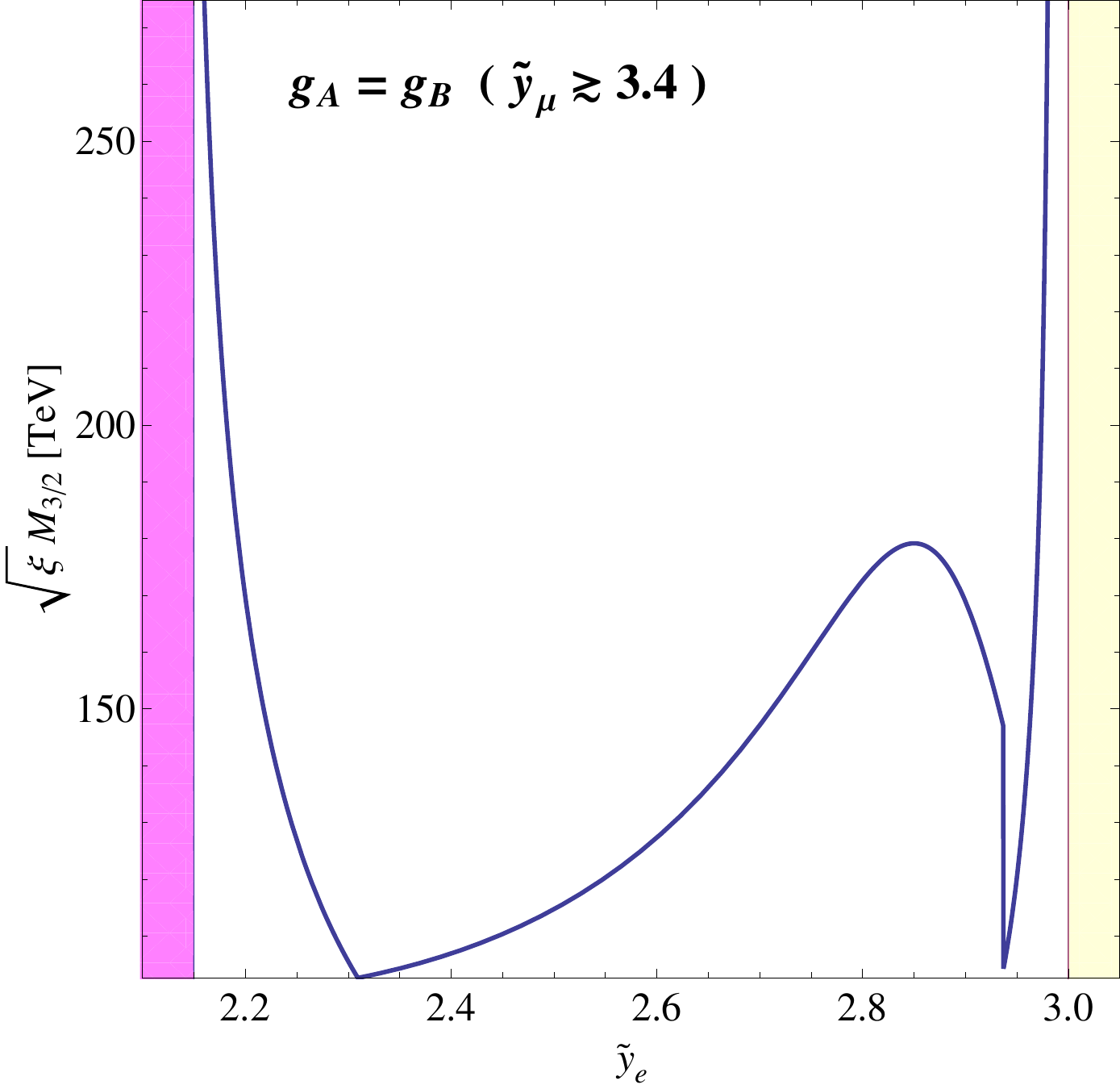}
\caption{{\em Left:} Contours of constant $\sqrt{\xi_f} \mgravitino$
as determined by the dark matter relic density in the $(\tilde{g}_B,
\tilde{y}_e)$ plane for fixed $\tilde{y}_\mu=5$.  The shaded regions
are excluded for the reasons indicated.  The green line segment at
$\tilde g_B=1$ indicates the domain for the plot in the right panel.
{\em Right:} $\sqrt{\xi_f} \mgravitino$ as a function of $\tilde{y}_e$
for fixed $\tilde{g}_B = 1$.  This curve is independent of
$\tilde{y}_\mu$, as long as $\tilde{y}_\mu \agt 3.4$, so that the
potential is stable for the entire $\tilde{y}_e$ range. Note the
cusp at $\tilde y_\mu\sim2.31$ and the discontinuity at
$\tilde y_\mu\sim2.94$, which
correspond to the dark matter making a transition from
one selectron mass eigenstate to another, and from a selectron to an
electron, respectively.  We have used the same shading as in the left panel to
indicate excluded regions.
\label{fig:relic}}
\end{figure}

For our U(1)$_A \times$U(1)$_B$ model, dark matter is either composed
of Dirac electrons annihilating to $A$-photons through $t$-channel
electrons, or selectrons $\tilde{e}_1$ annihilating to $A$-photons
through $t$-channel selectrons.  The annihilation constants are $k_e =
1$ for the electron and $k_{{\tilde e}_1} = 2$ for the
selectron~\cite{Feng:2008mu,Feng:2009mn}.  The resulting relic density
is
\begin{equation}
   \Omega_i \simeq 0.23 \, \xi_f \, \frac{1}{k_i}\,\left(\frac{0.025}{\alpha_A}
   \frac{m_i}{\tev}\right)^2
   = 0.23 \, \frac{f_i({\tilde g_B}, {\tilde y_e}, {\tilde y_\mu})}{k_i}
   \left( \frac{ \sqrt{\xi_f} \, \mgravitino}{126~\tev} \right)^2 \ ,
   \label{eq:relic}
\end{equation}
where $i=e$ or $\tilde{e}_1$, and we have defined the dimensionless
quantity
\begin{equation}
   f_i({\tilde g_B}, {\tilde y_e}, {\tilde y_\mu})
\equiv \frac{m_i^2}{M_{\tilde{A}}^2} \ ,
   \label{fidef}
\end{equation}
which depends only on the ratio of couplings.  The relic density is
therefore independent of the overall scale of the couplings, as
expected for WIMPless dark matter.  For every point in the parameter
space, $\sqrt{\xi_f} \mgravitino$ is fixed by the relic density.  In
\figref{relic}, $\sqrt{\xi_f} \mgravitino$ is plotted for the
$\tilde{y}_\mu=5$ and $\tilde{g}_B=1$ sections of the parameter space.

The gravitino mass in AMSB is bounded by colliders.  LEP2 constraints
require Wino masses $m_{\tilde{W}} > 92 - 103~\gev$, depending on the
chargino-neutralino mass difference~\cite{Feng:2009te}. Assuming the
minimal AMSB relation for the Wino mass, this implies $\mgravitino
\simeq 370 \, m_{\tilde{W}} \agt 34-38~\tev$.  The LHC also bounds the
gravitino mass, but these constraints depend on the spectrum of
strongly-interacting superpartners.  As an example, in the framework
of minimal AMSB~\cite{Gherghetta:1999sw,Feng:1999hg}, where a
universal scalar mass $m_0$ is added to solve the tachyonic slepton
problem, null results from the 0-lepton search by
ATLAS~\cite{daCosta:2011qk} imply $\mgravitino \agt 30-40~\tev$,
depending on the value of $\m_0$~\cite{Allanach:2011qr}.  These bounds
are also presumably relaxed if $R$-parity is violated, a viable
possibility, since the stability of dark matter does not require
$R$-parity conservation in this model.

{}From a low-energy phenomenological approach, a 40 TeV gravitino
would seem most natural.  Moreover, cosmological considerations lead
us to expect $\xi_f \sim 1$, which would result, for example, from the
case where the hidden and visible sectors were in thermal contact at
early times.  This points toward $\sqrt{\xi_f} \mgravitino \sim
\ord(100~\tev)$.  \Figref{relic} shows that such values are typical in
this model, and the desired thermal relic density is generically
obtained, as expected for a realization of the WIMPless miracle.

\section{Effects from New Relativistic Degrees of Freedom}
\label{sec:gstar}

\subsection{$g_*$ and $\xi$ at Freeze Out}

As was pointed out earlier, our model introduces several massless
particles.  Their existence may be used for estimating the value of
$\xi_f$ in \eqref{eq:relic}.  To see this, define $g_*(T)$ to be the
number of relativistic degrees of freedom at temperature $T$.
Assuming entropy conservation, the ratio of temperatures at freeze out
is given by
\begin{equation}
\xi_f = \left[ \frac{g_*^{\text{h}}(T_{\infty}^{\text{h}})}
{g_*^{\text{h}}(T_f^{\text{h}})} \,
\frac{g_*^{\text{v}}(T_f^{\text{v}})}
{g_*^{\text{v}}(T_{\infty}^{\text{v}})} \right]^{\frac{1}{3}}
\xi_{\infty} \ ,
\label{xi}
\end{equation}
where $\xi_\infty$ is the temperature ratio of the hidden and visible
sectors at very early (and very hot) times, and the superscripts ``h''
and ``v'' denote hidden and visible sector quantities, respectively.
In full generality, the value of $\xi_f$ depends on the field content
at all possible scales in both sectors.  However, assuming there are
no particles with masses between the temperature at which the two
sectors thermally decoupled and the masses of the heaviest particles
we have considered, we have $g_*^{\text{v}} (T_{\infty}^{\text{v}}) =
g_*^{\text{MSSM}} = 228.75$.  For the hidden sector we have
\begin{equation}
g_*^{\text{h}}(T_{\infty}^{\text{h}}) = \frac{7}{8}\left(2\times6 +
2 \times 2 \right) + \left( 2 \times 6 + 2 \times 2 \right )  = 30 \ .
\end{equation}
At the time of freeze out, the massless degrees of freedom in the
hidden sector are the photon, the Higgsino $\tilde{H}_{\mu}$, and the
muons, yielding
\begin{equation}
g_*^{\text{h}}(T_f^{\text{h}}) = \frac{7}{8}\left(4 + 2 \right) +
\left( 2  \right)  = \frac{29}{4} \ .
\end{equation}
\Eqref{xi} then gives
\begin{equation}
\xi_f = 1.25 \left[
 \frac{g_*^{\text{v}}(T_f^{\text{v}})}{106.75}
 \right]^{\frac{1}{3}} \xi_{\infty} \ ,
\label{xif}
\end{equation}
where we have normalized $g_*^{\text{v}}(T_f^{\text{v}})$ to the total
SM degrees of freedom $g_*^{\text{SM}} = 106.75$.  Assuming thermal
contact at early times ($\xi_\infty=1$), the value of $\xi_f$ remains
close to 1, which makes it easy to re-interpret the contours in
\figref{relic} as curves of constant $\mgravitino$.  Recall that the
lower bound from LHC is $\mgravitino \agt 30-40~\tev$.  Note, however,
that \eqref{xif} relies on the assumption of a ``high energy desert,''
as discussed above.  Moreover, light dark matter would imply lower
$g_*^{\text{v}}(T_f^{\text{v}})$ values, thereby decreasing
$\xi_f/\xi_\infty$.

\subsection{Bounds from CMB and BBN}

The massless particles of the hidden sector contribute to the number
of relativistic degrees of freedom at any temperature.  Their
existence is therefore constrained by the standard theory of BBN and
by observations of the CMB.  It is customary to measure the number of
extra degrees of freedom in units of the effective number of extra
neutrinos $\Delta N_{\text{eff}}$, as if these were new active
neutrino species contributing to the energy density of the universe.
Currently, some of the more stringent bounds on $\Delta\neff$ are
\begin{eqnarray}
\Delta \neff &=& 0.19 \pm 1.2 \ \text{(95\% CL)
BBN~\cite{Cyburt:2004yc,Fields:2006ga}} \ , \\
\Delta \neff &=& 1.51 \pm 0.75\ \text{(68\% CL)
CMB (ACT)~\cite{Dunkley:2010ge}} \ , \\
\Delta \neff &=& 0.81 \pm 0.42\ \text{(68\% CL)
CMB (SPT)~\cite{Keisler:2011aw}} \ ,
\end{eqnarray}
where the BBN constraint assumes a baryon density that has been fixed
to the value determined by the CMB, and both $^4$He and D data are
included, and the CMB constraints combine data from the indicated
experiments with WMAP 7-year results~\cite{Komatsu:2010fb}, distance
information from baryon acoustic oscillations, and Hubble constant
measurements.  The BBN result is fully consistent with the standard
model, but with relatively large uncertainty, while the CMB results
have smaller uncertainties and show 2$\sigma$ excesses.  In the near
future, the uncertainty in the measurement by Planck is expected to
drop to $\sim0.3$~\cite{Hamann:2007sb,Ichikawa:2008pz,Colombo:2008ta,%
Joudaki:2011nw}, given only $\sim 1$ year of data.  This should
improve further as soon as more data is acquired, and a future
LSST-like survey may determine $\Delta\neff$ with an accuracy within
$0.1$~\cite{Joudaki:2011nw}.  The current status of $\Delta \neff$ has
generated a great deal of interest; for recent reviews and possible
explanations, see, for example,
Refs.~\cite{Hamann:2011ge,Menestrina:2011mz}.

In the present context, we can express $\Delta\neff$ in terms of
$g_*^{\text{h}}$ and the temperature:
\begin{equation}
\Delta \neff \ \frac{7}{8} \ 2 \ T_{\nu}^4 = g_*^{\text{h}}
(T_{\text{CMB}}^{\text{h}}) T_{\text{CMB}}^{\text{h}\, 4} \ ,
\end{equation}
where $T_{\nu} = (4/11)^{1/3} T_{\text{CMB}}^{\text{v}}$.
Assuming entropy conservation, the values of $g_*$ at freeze out and
as measured by the CMB are related through
\begin{equation}
   \xi_{\text{CMB}}
   =\left[ \frac{g_*^{\text{h}}(T_f^{\text{h}})}
   {g_*^{\text{h}}(T_{\text{CMB}}^{\text{h}})}
   \frac{g_*^{\text{v}}(T_{\text{CMB}}^{\text{v}})}
   {g_*^{\text{v}}(T_f^{\text{v}})} \right]^{\frac{1}{3}} \xi_f \ .
   \label{gstarCMB}
\end{equation}
Using this relation, we get
\begin{eqnarray}
\Delta \neff &=& \frac{4}{7} \left( \frac{11}{4} \right)^{\frac{4}{3}}
g_*^{\text{h}} (T_{\text{CMB}}^{\text{h}}) \,
\xi_{\text{CMB}}^4 \\
&=& \frac{4}{7} \left( \frac{11}{4} \right)^{\frac{4}{3}}
g_*^{\text{h}} (T_{\text{CMB}}^{\text{h}}) \,
\left[ \frac{g_*^{\text{h}}(T_f^{\text{h}})}
{g_*^{\text{h}}(T_{\text{CMB}}^{\text{h}})}
\frac{g_*^{\text{v}}(T_{\text{CMB}}^{\text{v}})}
{g_*^{\text{v}}(T_f^{\text{v}})} \right]^{\frac{4}{3}} \xi_f^4 \ .
\label{Nnu}
\end{eqnarray}
At the time of CMB decoupling we have
$g_*^{\text{v}}(T_{\text{CMB}}^{\text{v}})=2$ and
$g_*^{\text{h}}(T_f^{\text{h}}) =
g_*^{\text{h}}(T_{\text{CMB}}^{\text{h}}) =29/4$.  This implies
\begin{equation}
   \Delta \neff = \left(\frac{\xi_f}{1.88} \right)^4
   \left[ \frac{106.75}
   {g_*^{\text{v}}(T_f^{\text{v}})} \right]^{\frac{4}{3}} \ .
   \label{dneffxi}
\end{equation}
We may use now \eqref{xif} to express the effective number of extra
neutrinos in terms of $\xi_{\infty}$.  Under the assumption of a high
energy desert we obtain
\begin{equation}
\Delta \neff = 0.19 \, \xi_{\infty}^4 \ .
\label{dneff}
\end{equation}
Moreover, note that \eqref{dneff} is independent of
$g_*^{\text{v}}(T_f^{\text{v}})$, giving a sharp prediction once the
two assumptions of a high energy desert and thermal contact at early
times ($\xi_\infty=1$) are made.  Such a prediction is interesting,
especially given the bright prospects for improved measurements of
$\Delta \neff$ in the near future.

\begin{figure}
  \centering
    \includegraphics[width=0.50\textwidth]{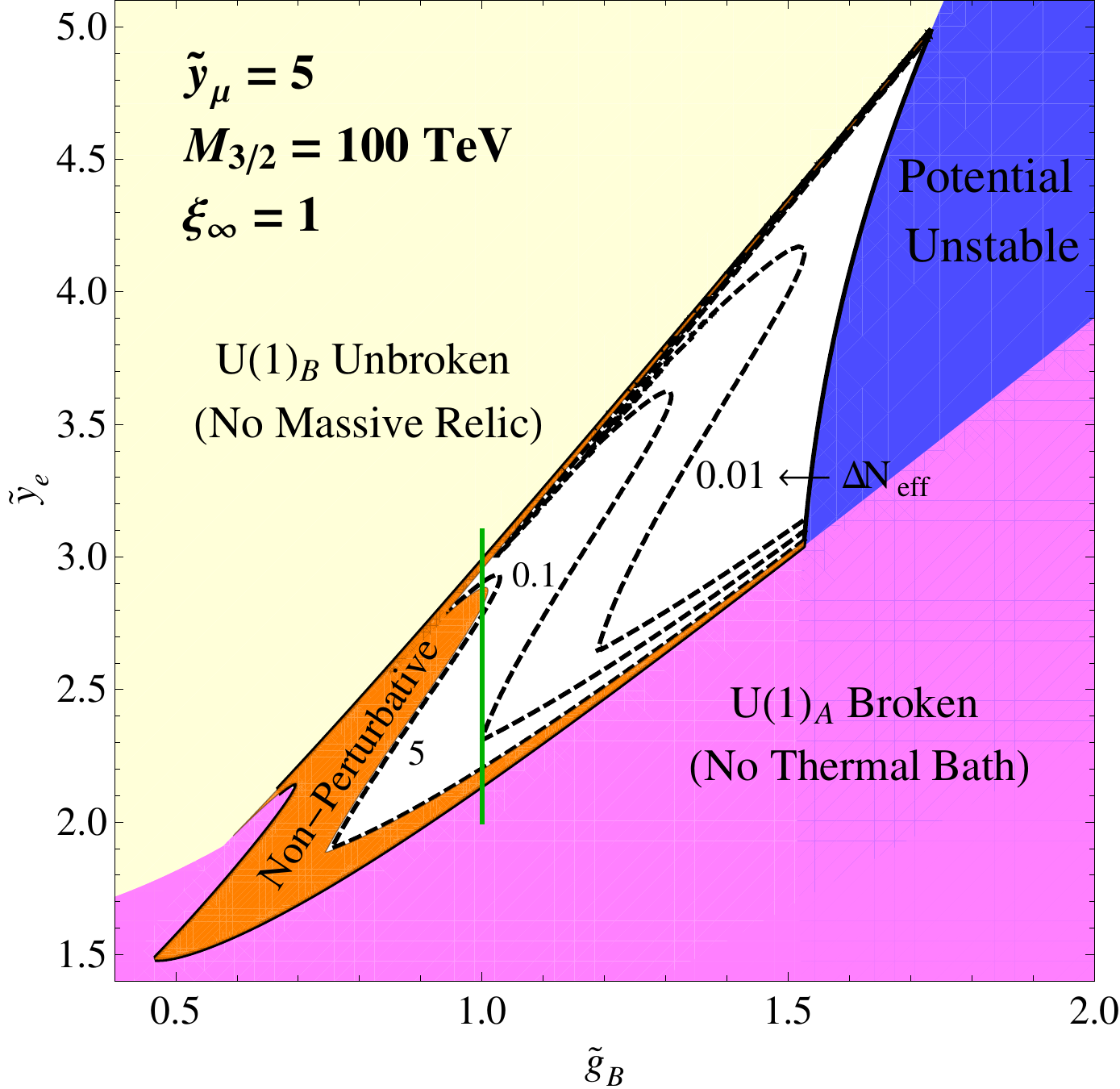} \hfil
    \includegraphics[width=0.48\textwidth]{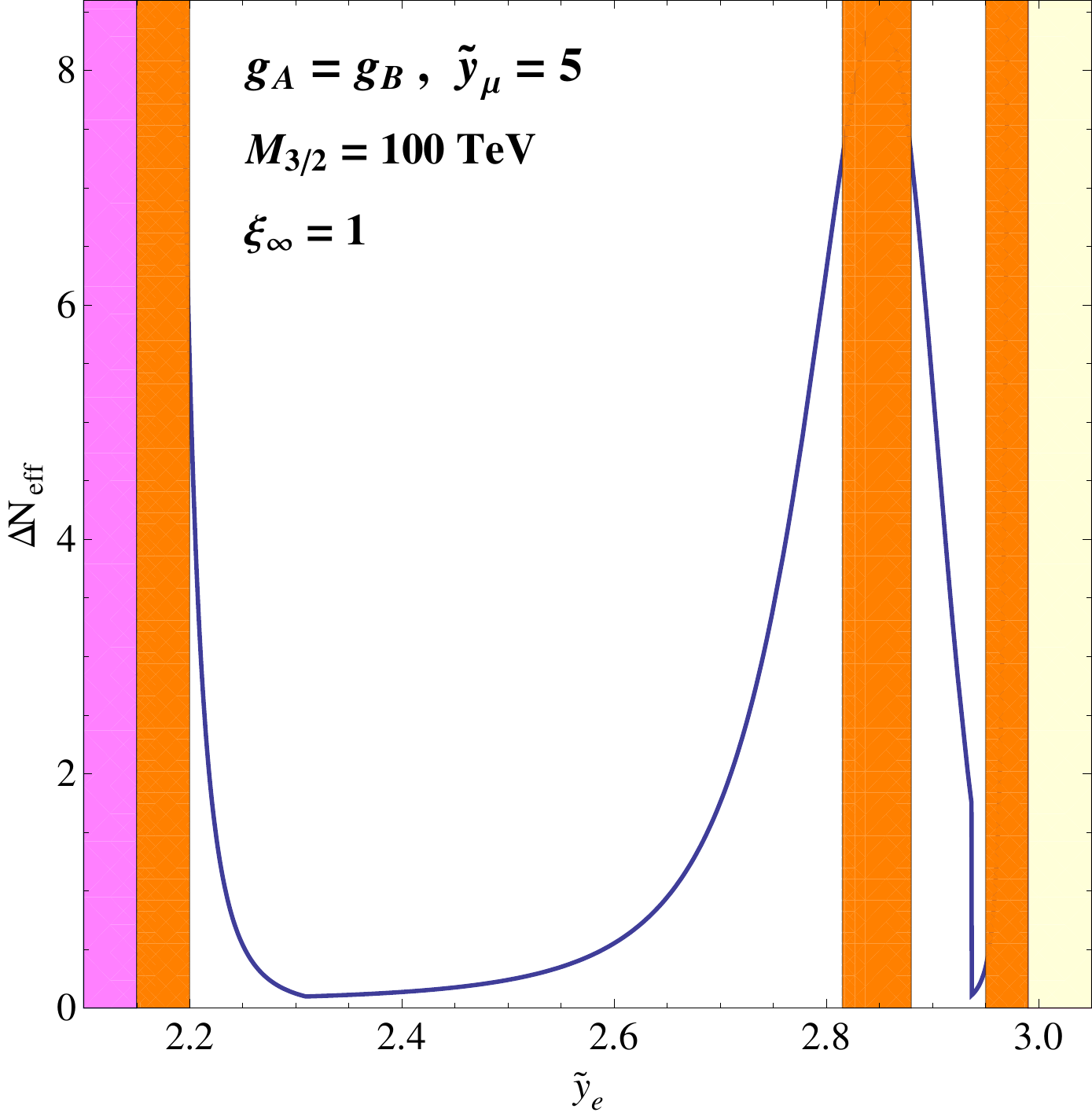}
\caption{{\em Left:} Contours of constant $\Delta\neff$ in the
$(\tilde{g}_B, \tilde{y}_e)$ plane for fixed $\tilde{y}_\mu=5$,
$\mgravitino=100~\tev$, and $\xi_{\infty} = 1$.  The shaded regions
are excluded for the reasons indicated.  The orange region, labeled
``Non-Perturbative,'' is excluded by considerations of
self-interactions and perturbativity, as explained in
\secref{selfinteractions}.  The green line segment at $g_A=g_B$ indicates the
domain taken for the plot in the right panel.  {\em Right:}
$\Delta\neff$ as a function of $\tilde{y}_e$ for the same parameters
as in the left panel and $\tilde{g}_B = 1$.  This curve is independent
of $\tilde{y}_\mu$, as long as $\tilde{y}_\mu \agt 3.4$, so that the
potential is stable for the entire $\tilde{y}_e$ range.  We have used
the same shading as in the left panel to indicate excluded regions.
\label{fig:neff}}
\end{figure}

Alternatively, given $\mgravitino$, we can obtain $\xi_f$ as a
function of the parameter space, as determined by the relic density
condition.  This implies, through \eqref{dneffxi}, that $\Delta\neff$
is determined as well.  In \figref{neff}, $\Delta\neff$ is plotted for
the sections of parameter space defined by $\tilde{y}_\mu=5$ and
$\tilde{g}_B= 1$.  Note, however, that $\Delta \neff$ is highly
sensitive to $\mgravitino$: for a fixed relic density, $\Delta \neff
\propto \mgravitino^{-8}$.

\section{Self-Interactions}
\label{sec:selfinteractions}

So far, all the observables we have discussed depend only on ratios of
couplings.  This scaling is a key feature of WIMPless dark matter.
However, some observations constrain absolute coupling values, rather
than just ratios.

An example is constraints from structure formation. The dark matter
described in this work has a hidden charge, and is therefore subject
to constraints on self-interactions through a long-range force.  In
Refs.~\cite{Ackerman:2008gi,Feng:2009mn}, bounds on dark matter mass
and coupling were derived from the observation of elliptical halos.
Following earlier work~\cite{2002ApJ...564...60M}, the authors used
measurements that established the ellipticity of the galaxy NGC
720~\cite{Buote:2002wd,Humphrey:2006rv}.  Strong enough
self-interactions would tend to turn elliptic halos into spheres over
the course of a cosmological time scale, leading to the bound
\begin{equation}
   \lt(\frac{m_X}{22\,\tev}\rt)^3 \gtrsim \alpha_X^2 \ .
\end{equation}
Using $\alpha_A = \pi M_{\tilde{A}} / \mgravitino$ and \eqref{fidef},
we obtain the lower bound
\begin{equation}
   m_i \agt  \frac{10~\tev}{f_i\lt(\tilde g_B,\tilde y_e,\tilde y_\mu\rt)}
   \lt(\frac{100\,\tev}{\mgravitino}\rt)^2\equiv m_{\text{DM}}^{\text{min}}\ ,
   \label{eq:mDMmin}
\end{equation}
where $i$ denotes either $e$ or $\tilde{e}_1$, depending on the
identity of the dark matter particle at the particular point of
parameter space.

\begin{figure}
\centering
\includegraphics[width=0.483\textwidth]{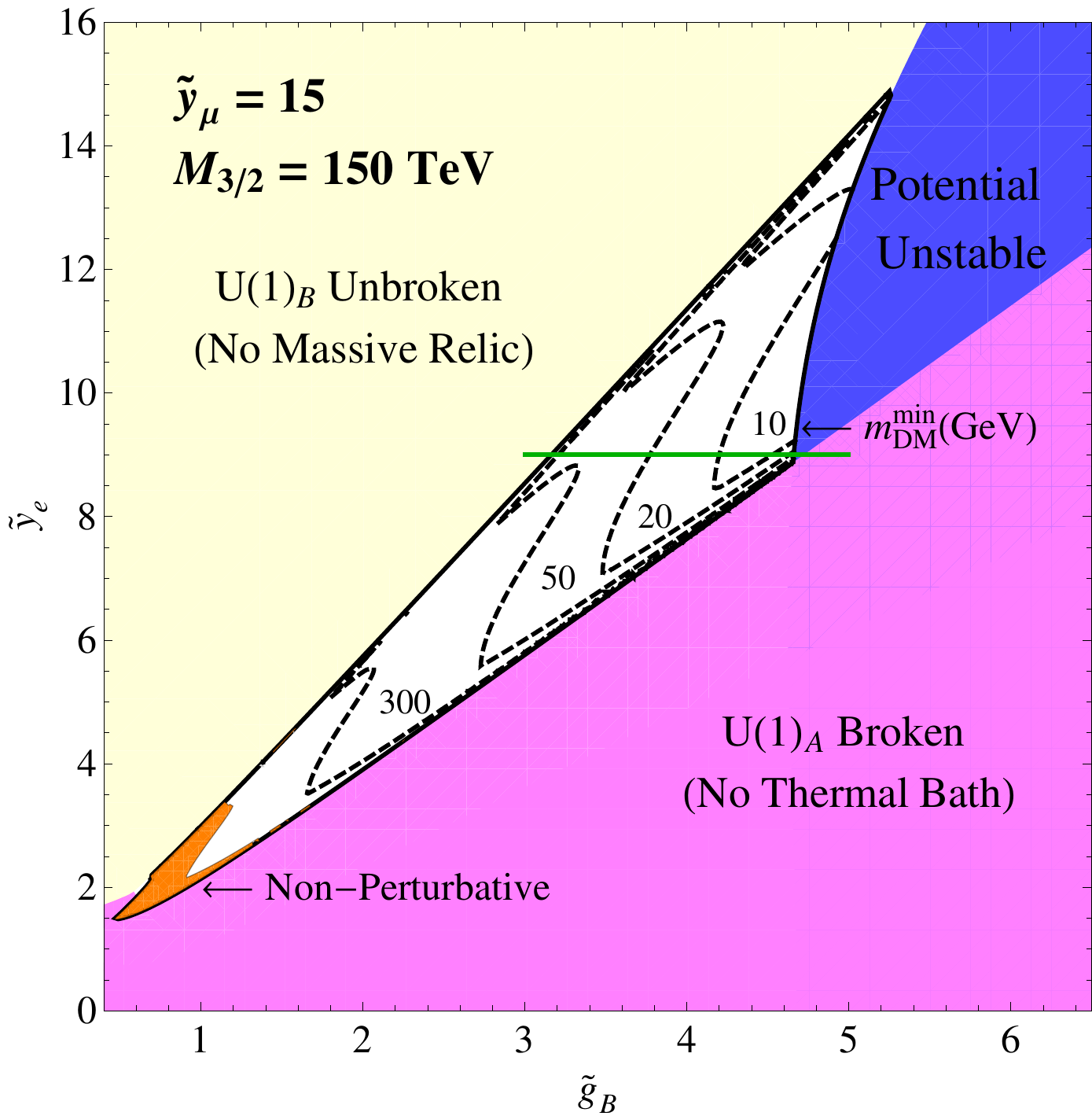} \hfil
\includegraphics[width=0.497\textwidth]{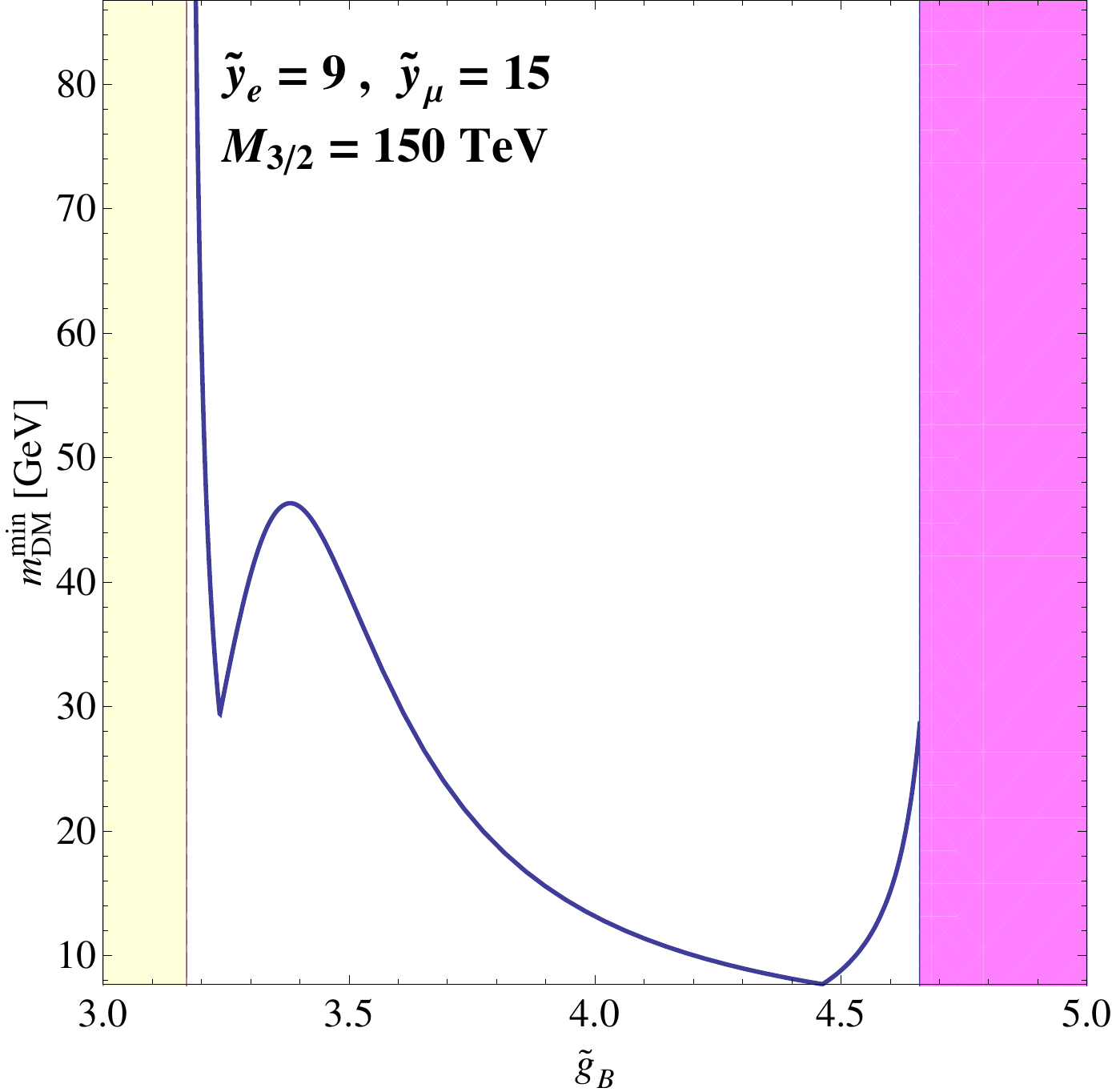}
\caption{{\em Left:} Contours of constant $m_{\text{DM}}^{\text{min}}$
in the $(\tilde{g}_B, \tilde{y}_e)$ plane for fixed $\tilde{y}_\mu =
15$ and $\mgravitino=150~\tev$.  The shaded regions are excluded for
the reasons indicated.  The green line segment at $\tilde y_e=9$
indicates the domain
taken for the plot in the right panel.  {\em Right:} The minimum dark
matter mass $m_{\text{DM}}^{\text{min}}$ as a function of
$\tilde{g}_B$ for the same parameters as in the left panel and
$\tilde{y}_e = 9$.  This curve is independent of $\tilde{y}_\mu$, as
long as $\tilde{y}_\mu \agt 3.4$, so that the potential is stable for
the entire $\tilde{y}_e$ range.  We have used the same shading as in
the left panel to indicate excluded regions.
\label{fig:mDMmin}}
\end{figure}

This lower bound on the dark matter mass also sets a lower bound on
the mass of the heaviest particle in the spectrum at each point in the
parameter space. However, our description above relies on a
perturbative expansion that is valid as long as all particle masses
(and in particular the heaviest particle mass) are below
$\mgravitino$~\cite{Feng:2011uf}. As a result, certain regions in the
parameter space are excluded for a given $\mgravitino$.
\Figref{mDMmin} shows contours of constant
$m_{\text{DM}}^{\text{min}}$ according to \eqref{eq:mDMmin}.  Regions
that are forbidden by perturbativity (or breakdown of the effective
field theory) are shown as well.  As can be seen in the figure, dark
matter can be as light as a few GeV for reasonable values of
$\mgravitino$ and $\tilde y_\mu$.  Smaller dark matter masses are also
possible if one tunes parameters to more extreme values.  Values of
dark matter mass $\sim 10~\gev$ are of special interest, given
reported direct detection signals of dark matter with such masses.  Of
course, a complete explanation of such signals requires coupling the
hidden sector to the visible sector, which we have not done in this paper.

\section{Summary}
\label{sec:summary}

In this work, we have presented a model for WIMPless dark matter from
a hidden sector with AMSB.  The novel feature of this work is that
dark matter in a hidden sector naturally has the correct relic
density, in the sense that it is determined purely by the soft SUSY
breaking scale, without the introduction and tuning of other
dimensionful parameters.  The correct relic density therefore emerges
naturally, in the same sense as for WIMPs, but the dark matter may
have very different masses and interaction strengths.

Our new model has a U(1)$\times$U(1) gauge symmetry. One U(1) provides
massless hidden photons for the thermal bath, and the second U(1) is
broken spontaneously by a Higgs field.  The matter field content
includes a family of three chiral superfields, and its mirror family,
with all the charges inverted.  The mirror family is required for the
cancellation of chiral anomalies, but we prevent renormalizable
supersymmetric inter-family couplings by imposing a $Z_3$ symmetry,
such that all the fields have the same triality.  Symmetries therefore
forbid the introduction of new mass scales.  The symmetries also
guarantee the stability of a massive dark matter candidate.
$R$-parity conservation is not required, and so the visible sector may
appear at colliders through $R$-parity violating signals.  We note,
however, that since the $Z_3$ symmetry is spontaneously broken, the
model suffers from domain wall problems, similar to those of the
next-to-minimal supersymmetric standard model.  We assume that these
may be overcome through similar mechanisms (for discussions, see, for
example, Refs.~\cite{Preskill:1991kd,Abel:1995wk,Ellwanger:2009dp}),
but a detailed investigation is beyond the scope of this work.

The dark matter spectrum depends on two gauge and two Yukawa
couplings, while annihilation depends exclusively on the gauge
coupling of the unbroken U(1).  However, the relic density depends
only on ratios of couplings, and not on their overall scale.  For
non-hierarchical couplings, the correct relic density is obtained,
irrespective of the dark matter's mass or interaction strength,
thereby realizing the WIMPless miracle.

The model includes new relativistic degrees of freedom contributing to
the energy density of the universe at freeze out and at late times.
In a significant region of the parameter space, non-zero values of
$\Delta\neff$ are predicted.  This observable is now being probed by
the Planck observatory.  This model also predicts dark matter that
self-interacts through long-range interactions.  Such
self-interactions are constrained by halo shapes, but provide another
possible astrophysical signal for this dark matter scenario (and
others).

The self-interaction bounds also impose a lower bound on the dark
matter mass.  Nevertheless, regions in the parameter space where this
bound is low (for example, below 10 GeV for $\tilde y_\mu=15$ and
$\mgravitino=150~\tev$) are allowed.  It would be interesting to
relieve the stringent constraints imposed by galactic halo shapes by
giving the hidden photon a small mass, or to couple the hidden and
visible sectors to each other through the kinetic mixing of hidden and
visible photons.  In such a scenario, the hidden photon could decay to
the SM, potentially giving rise to interesting collider and dark
matter detection phenomenology.  Such possibilities are of special
interest, given that this scenario provides dark matter with the
correct thermal relic density that is nevertheless light, as may be
indicated by current signals in direct detection experiments.

\section*{Acknowledgments}
We thank Manoj Kaplinghat, Yuri Shirman, and Yael Shadmi for helpful
conversations.  The work of JLF was supported in part by NSF grant
PHY--0970173. The work of VR was supported in part by DOE grant
DE-FG02-04ER-41298.

\providecommand{\href}[2]{#2}\begingroup\raggedright\endgroup

\end{document}